# REGULATING RESPONSE TIME IN AN AUTONOMIC COMPUTING SYSTEM: A COMPARISON OF PROPORTIONAL CONTROL AND FUZZY CONTROL APPROACHES

Harish S. Venkatarama[1] and Kandasamy Chandra Sekaran[2]

[1]Reader, Computer Science & Engg. Dept.,
Manipal Institute of Technology, Manipal, India
harish.sv@manipal.edu

[2]Professor, Dept. of Computer Engg.,
National Institute of Technology Karnataka, India
kch@nitk.ac.in

## ABSTRACT

*Ecommerce is an area where an Autonomic Computing system could be very effectively deployed. Ecommerce has created demand for high quality information technology services and businesses are seeking quality of service guarantees from their service providers. These guarantees are expressed as part of service level agreements. Properly adjusting tuning parameters for enforcement of the service level agreement is time-consuming and skills-intensive. Moreover, in case of changes to the workload, the setting of the parameters may no longer be optimum. In an ecommerce system, where the workload changes frequently, there is a need to update the parameters at regular intervals. This paper describes two approaches, one, using a proportional controller and two, using a fuzzy controller, to automate the tuning of MaxClients parameter of Apache web server based on the required response time and the current workload. This is an illustration of the self-optimizing characteristic of an autonomic computing system.*

## 1. INTRODUCTION

The advent and evolution of networks and Internet, which has delivered ubiquitous service with extensive scalability and flexibility, continues to make computing environments more complex [1]. Along with this, systems are becoming much more software-intensive, adding to the complexity. There is the complexity of business domains to be analyzed, and the complexity of designing, implementing, maintaining and managing the target system. I/T organizations face severe challenges in managing complexity due to cost, time and relying on human experts. All these issues have necessitated the investigation of a new paradigm, Autonomic computing [1], to design, develop, deploy and manage systems by taking inspiration from strategies used by biological systems. Ecommerce is one area where an Autonomic Computing system could be very effectively deployed. Ecommerce has created demand for high quality information technology (IT) services and businesses are seeking quality of service (QoS) guarantees from their service providers (SPs). These guarantees are expressed as part of service level agreements (SLAs). As an example, performance of an Apache web server [16] is heavily influenced by the MaxClients parameter, but the optimum value of the parameter depends on system capacity, workload and the SLA. Properly adjusting tuning parameters for enforcement of the SLA is time-consuming and skills-intensive. Moreover, in case of changes to the workload, the setting of the parameters may no longer be optimum. In an ecommerce system, where the workload changes frequently, there is a need to update the parameters at regular intervals.

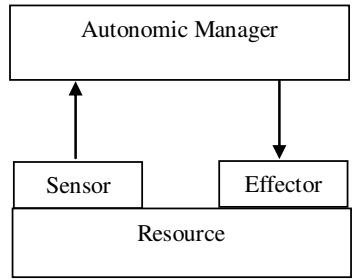

Figure 1. Autonomic computing architecture

The simplified architecture for autonomic computing is shown in figure 1. Adding an autonomic manager makes the resource self-managing [2]. The manager gets required data through the sensors and regulates the behavior of the resource through effectors. This shows how self-managing systems are developed using feedback control loops. This observation suggests that control theory will be of help in the construction of autonomic managers.

Control theory has been applied to many computing systems, such as networks, operating systems, database management systems, etc. The authors in [3] propose to control web server load via content adaptation. The authors in [5] extend the scheme in [3] to provide performance isolation, service differentiation, excess capability sharing and QoS guarantees. In [4][8] the authors propose a relative differentiated caching services model that achieves differentiation of cache hit rates between different classes. The same objective is achieved in [6], which demonstrates an adaptive control methodology for constructing a QoS-aware proxy cache. The authors in [7] present the design and implementation of an adaptive architecture to provide relative delay guarantees for different service classes on web servers.

Real-time scheduling theory makes response-time guarantees possible, if server utilization is maintained below a pre-computed bound. Feedback control is used in [9] to maintain the utilization around the bound. The authors in [10][11] demonstrate the power of a control theoretic analysis on a controller for doing admission control of a Lotus Notes workgroup server.

MIMO techniques are used in [12][13] to control the CPU and memory utilization in web servers. Queuing theory is used in [14] for computing the service rate necessary to achieve a specified average delay given the currently observed average request arrival rate. Same approach is used to solve the problem of meeting relative delay guarantees in [15].

The authors in [18] present a framework that monitors client perceived service quality in real-time with considerations of both network transfer time and server-side queuing delays and processing time. The authors in [19] present a fuzzy controller to guarantee absolute delays.

The authors in [20] present a Linear-Parameter-Varying approach to the modeling & design of admission control for Internet web servers. The authors in [21] [22] study the performance/power management of a server system.

The authors in [23] propose an approach to automate enforcement of SLAs by constructing IT level feedback loops that achieve business objectives, especially maximizing SLA profits (the difference between revenue and costs). Similarly, the authors in [24] propose a profit-oriented feedback control system that automates the admission control decisions in a way that balances the loss of revenue due to rejected work against the penalties incurred if admitted work has excessive response times. The authors in [25] describe an approach to automate parameter tuning using a fuzzy controller that employs rules incorporating qualitative knowledge of the effect of tuning parameters.

This paper targets two objectives. Initially an approach to automate the tuning of MaxClients parameter of Apache web server using a proportional controller is explained. As a second objective, an approach to automate the tuning of MaxClients parameter of Apache web server using a fuzzy controller is described. In both the cases the controller maximizes the number of users allowed to connect to the system subject to the response time constraint as given in the SLA. This is an illustration of the self-optimizing characteristic of an autonomic computing system.

## 2. SYSTEM BACKGROUND

The system studied here is the Apache web server. In Apache version 2.2 (configured to use Multi-Processing Module prefork), there are a number of worker processes monitored and controlled by a master process [16]. The worker processes are responsible for handling the communications with the web clients. A worker process handles at most one connection at a time, and it continues to handle only that connection until the connection is terminated. Thus the worker is idle between consecutive requests from its connected client.

A parameter termed MaxClients limits the size of this worker pool, thereby providing a kind of admission control in which pending requests are kept in the queue. MaxClients should be large enough so that more clients can be served simultaneously, but not so large that response time constraints are violated. If MaxClients is too small, there is a long delay due to waits in the queue. If it is too large resources become over utilized which degrades performance as well. The optimal value depends on server capacity, nature of the workload and the SLA.

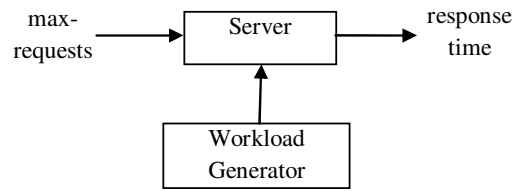

Figure 2. Modelling the system

## 3. MODELING AND SYSTEM IDENTIFICATION

Any conventional controller design starts with modeling and system identification. Figure 2 shows the scheme used for this purpose. The simulation environment consists of a workload generator which generates requests and a server program which services the requests.

In this model, parameter max-requests is varied from 200 in steps of 10 and the corresponding response time values are noted. A first order ARX model is used to describe the relationship between inputs and outputs.

$$y(k+1) = a*y(k) + b*u(k) \qquad (1)$$

Here, u is the input or actuating signal, y is the output signal and a and b are scalars. Since a discrete signal has value only at specific instants of time, an integer k is used to index these instants. Using least squares regression, values for a and b are estimated as a = 0.1 and b = -0.36. That is, we arrive at the model

$$y(k+1) = 0.1*y(k) - 0.36*u(k) \qquad (2)$$

## 4. PROPORTIONAL CONTROLLER DESIGN AND IMPLEMENTATION

We use the proportional control law.

$$u(k) = K_P*e(k) \qquad (3)$$

Here $K_P$ is a constant called gain of the controller. The actuating signal is proportional to the present error signal. It is not dependent on the past values of the error. Taking Z transform of equation (2) and manipulating, we get the open loop transfer function.

$$G(z) = Y(z) / U(z) = -0.36 / (z-0.1) \qquad (4)$$

Closed loop transfer function is as follows.

$$F_R(z) = Y(z) / R(z) = K_P*G(z) / (1 + K_P*G(z))$$

Solution of the characteristic equation, $1 + K_P*G(z) = 0$ gives the poles. For the system in question, there is only 1 closed loop pole, given by the following equation.

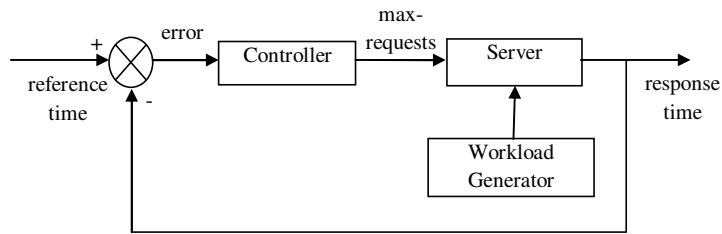

Figure 3. System for Proportional control

$$p1 = 0.1 + 0.36*K_P$$

For stability, we need to have, $| 0.1 + 0.36*K_P | < 1$ or $-3.1 < K_P < 2.5$

Figure 3 shows the system for proportional control which is used for the implementation. In terms of figure 1, server is the resource and controller is the autonomic manager. Response time is converted to error signal, which corresponds to input to the manager from the sensor. Just as the behavior of the resource is influenced by the effector, the server is influenced by max-requests.

The incoming request from the workload generator is first put into a queue in the server. When the server becomes free, the first request in the queue is dequeued. The time spent by the request in the queue is called the response time. The workload generator generates requests such that the time between generations of consecutive requests is exponentially distributed. Also, the time taken by the server to process each request is exponentially distributed. Thus, the client server architecture is simulated here as an M/M/1 queue.

Workload generator is set to generate requests such that the time between arrivals of consecutive requests on an average (mean interarrival) is 0.2 second. That is 300 requests per minute on an average. Mean service time is set to 60 seconds. Readings are noted every 3 minutes. To ensure that transients do not affect the readings, readings are taken for the last 1 minute of the 3 minute interval. Response time values of the requests which entered service in the last 1 minute are noted and the average is calculated. In this simulation, MaxClients is simulated by max-requests. Gain $K_P$ is set to -1.5.

The controller tries to drive the error signal to 0 by adjusting the value of max-requests at regular intervals. That is, it tries to make the response time equal to the reference time. The simulation is carried out using C-based simulation language "simlib" [17]. Each simulation was run for 60 minutes.

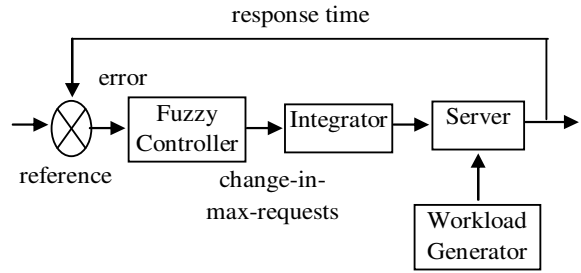

Figure 4.  Block diagram of the fuzzy control system

## 5. FUZZY CONTROLLER DESIGN AND IMPLEMENTATION

The block diagram of the fuzzy control system is shown in figure 4. The simulation environment consists of a workload generator program to generate requests, a server program to service the requests, a fuzzy controller program and an integrator routine.

The incoming request from the workload generator is first put into a queue in the server. When the server becomes free, the first request in the queue is dequeued. The time spent by the request in the queue is called the response time. Here also, the client server architecture is simulated as an M/M/1 queue. The number of requests accepted by the server is limited by the parameter max-requests, which is updated by the integrator at the beginning of every measurement interval. Simulation readings are recorded after every interval, called measurement interval.

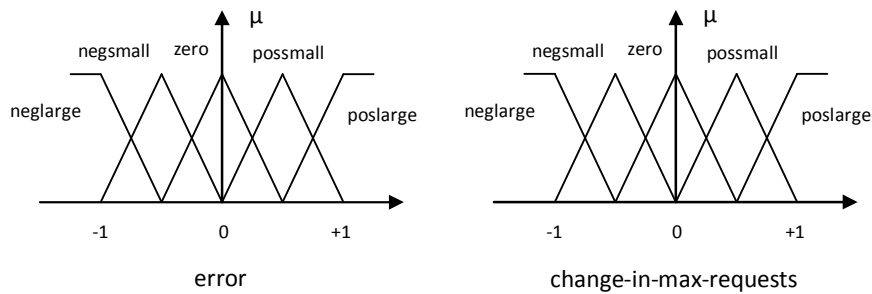

Figure 5.  Membership functions

Table 1. Fuzzy Rules

| Rule | IF error | THEN change-in-max-requests |
|---|---|---|
| 1 | neglarge | poslarge |
| 2 | negsmall | possmall |
| 3 | zero | zero |
| 4 | possmall | negsmall |
| 5 | poslarge | neglarge |

Any fuzzy control system involves three main steps, that is, fuzzification, inference mechanism and defuzzification [26]. Figure 5 shows the triangular membership functions used for the fuzzification of the input and defuzzification of the output. In each case, the parameter is divided into 5 intervals called neglarge, negsmall, zero, possmall and poslarge. Neglarge is an abbreviation for "negative large in size". Similarly negsmall, possmall and poslarge are abbreviations. Zero is the name of the interval denoting small changes. The measured numeric values will be multiplied by factors known as the normalized gains. That is why the x-axis shows -1 and 1 for all the membership functions. The output value, change-in-max-requests, obtained will be denormalized by dividing by the normalized gain to obtain the actual output value. The fuzzy rules describing the working of the controller is shown in table 1.

As before, workload generator is set to generate requests with mean interarrival equal to 0.2

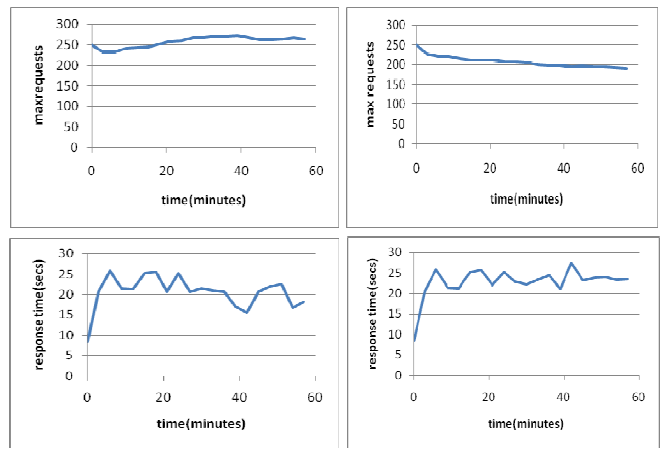

Figure 6. Results for proportional control with reference time
= 20 secs (left hand side) and 25 secs (right hand side)

second and mean service time is set to 60 seconds. The fuzzy controller program takes as input, error, which is response-time subtracted from the reference value. The controller calculates the adjustment required for max-requests, i.e., change-in-max-requests for the next measurement interval. This value is sent to the integrator, which calculates the value of max-requests for the next interval. The measurement interval and the method for noting the readings is the same as before.

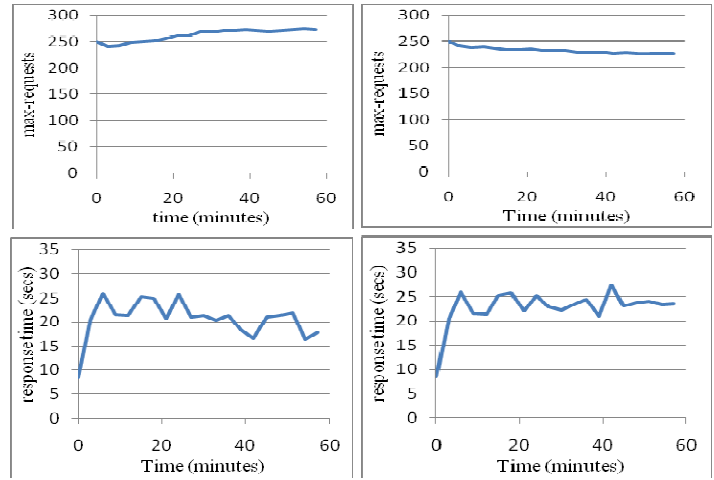

Figure 7. Results for fuzzy control with reference time
= 20 secs (left hand side) and 25 secs (right hand side)

## 6. RESULTS

The simulation was carried out for different values of reference times. Each simulation was run for 60 minutes. Figure 6 shows the results for the proportional controller for reference time = 20 seconds (top and bottom figures on left hand side) and 25 seconds (top and bottom figures on right hand side) respectively. The plots at the top of the figure show the variation of max-requests, while the plots at the bottom show the variation in response time.

Figure 7 shows the results for fuzzy controller for reference time = 20 seconds (top and bottom figures on left hand side) and 25 seconds (top and bottom figures on right hand side) respectively. The plots at the top of the figure show the variation of max-requests, while the plots at the bottom show the variation in response time.

There is not much difference in terms of performance of the controllers with respect to regulation of response time. Further, for higher value of reference time, smaller value of max-requests suffices. This is true in both cases. However, proportional control is seen to be more efficient, since the regulation is done by having comparatively smaller values of max-requests, which means the resource requirement is lesser. This is true irrespective of reference time. The disadvantage of proportional control is that the system to be controlled has to first modeled. Moreover, in case of changes to the system or workload, the model may no longer be valid. As seen, fuzzy controller design does not need modeling of the system. In this sense, it is independent of the model.

## 8. CONCLUSIONS

This paper describes two approaches to regulate response time in an ecommerce system. One approach uses proportional control while the other uses fuzzy control. Proportional control leads to smaller value of max-requests, but the downside is the modeling and system identification step. A proportional controller design is very closely tied to the system characteristics. If the parameters of the system are expected to be relatively constant, then a proportional controller may be a better choice, given its efficiency. Otherwise a fuzzy controller does a better job. So the choice of controllers, depends on the system to be controlled.

These are illustrations of the self-optimizing characteristic of an autonomic computing system. Specifically, the system studied here is tuning of MaxClients parameter of the Apache web server to satisfy the parameters mentioned in the SLA. The workload and server are simulated as an M/M/1 queue. The controller attempts to maximize max-requests, which is equivalent to MaxClients. It is easily seen from the results, that a single fixed value of max-requests will not be optimum for all cases. Since workload of a server can change rapidly, it is of immense benefit to have a controller which updates the value of MaxClients at regular intervals.

Though the controllers are properly able to adjust value of max-requests, it takes some time for them to converge to the optimum value. Thus, as part of future work, it is intended to find ways to speed up the working of the system. It is also intended to test the functioning of the controllers under different simulation environments like having an arbitrary (general) distribution for the service time, i.e., simulating the workload and server as an M/G/1 queue.